\documentclass[aps,prb,twocolumn,showpacs,superscriptaddress]{revtex4}

\usepackage{graphicx}
\usepackage{color}
\usepackage[normalem]{ulem}

\begin{document}

\preprint{ver. 3.0}

\title{Soft and Isotropic Phonons in PrFeAsO$_{1-y}$}


\author{T.~Fukuda}
\affiliation{Materials Dynamics Laboratory, RIKEN SPring-8 Center, Sayo,
Hyogo 679-5148, Japan}
\affiliation{Quantum Beam Science Directorate, JAEA SPring-8, Sayo,
Hyogo 679-5148, Japan}
\affiliation{JST, Transformative Research-Project on Iron Pnictides
(TRIP), Chiyoda, Tokyo, 102-0075, Japan}
\author{A.Q.R.~Baron}
\affiliation{Materials Dynamics Laboratory, RIKEN SPring-8 Center, Sayo,
Hyogo 679-5148, Japan}
\affiliation{Research and Utilization Division, JASRI SPring-8, Sayo,
Hyogo 679-5198, Japan}
\affiliation{JST, Transformative Research-Project on Iron Pnictides
(TRIP), Chiyoda, Tokyo, 102-0075, Japan}
\author{H.~Nakamura}
\affiliation{JST, CREST and the Center for Computational Science and
e-Systems, JAEA, Taito, Tokyo, 110-0015, Japan}
\affiliation{JST, Transformative Research-Project on Iron Pnictides
(TRIP), Chiyoda, Tokyo, 102-0075, Japan}
\author{S.~Shamoto}
\affiliation{Quantum Beam Science Directorate, JAEA, Naka, Ibaraki
319-1195, Japan}
\affiliation{JST, Transformative Research-Project on Iron Pnictides
(TRIP), Chiyoda, Tokyo, 102-0075, Japan}
\author{M.~Ishikado}
\affiliation{Quantum Beam Science Directorate, JAEA, Naka, Ibaraki
319-1195, Japan}
\affiliation{JST, Transformative Research-Project on Iron Pnictides
(TRIP), Chiyoda, Tokyo, 102-0075, Japan}
\author{M.~Machida}
\affiliation{JST, CREST and the Center for Computational Science and
e-Systems, JAEA, Taito, Tokyo, 110-0015, Japan}
\affiliation{JST, Transformative Research-Project on Iron Pnictides
(TRIP), Chiyoda, Tokyo, 102-0075, Japan}
\author{H.~Uchiyama}
\affiliation{Materials Dynamics Laboratory, RIKEN SPring-8 Center, Sayo,
Hyogo 679-5148, Japan}
\affiliation{Research and Utilization Division, JASRI SPring-8, Sayo,
Hyogo 679-5198, Japan}
\author{A.~Iyo}
\affiliation{Nanoelectronics Research Institute (NeRI), AIST, Tsukuba,
Ibaraki, 305-8568, Japan}
\affiliation{JST, Transformative Research-Project on Iron Pnictides
(TRIP), Chiyoda, Tokyo, 102-0075, Japan}
\author{H.~Kito}
\affiliation{Nanoelectronics Research Institute (NeRI), AIST, Tsukuba,
Ibaraki, 305-8568, Japan}
\affiliation{JST, Transformative Research-Project on Iron Pnictides
(TRIP), Chiyoda, Tokyo, 102-0075, Japan}
\author{J.~Mizuki}
\affiliation{Quantum Beam Science Directorate, JAEA SPring-8, Sayo,
Hyogo 679-5148, Japan}
\affiliation{JST, Transformative Research-Project on Iron Pnictides
(TRIP), Chiyoda, Tokyo, 102-0075, Japan}
\author{M.~Arai}
\affiliation{J-PARC Center, JAEA, Naka, Ibaraki, 319-1195, Japan}
\author{H.~Eisaki}
\affiliation{Nanoelectronics Research Institute (NeRI), AIST, Tsukuba,
Ibaraki, 305-8568, Japan}
\affiliation{JST, Transformative Research-Project on Iron Pnictides
(TRIP), Chiyoda, Tokyo, 102-0075, Japan}


\date{\today}

\begin{abstract}
 Phonons in single crystals of PrFeAsO$_{1-y}$ are investigated using
 high-resolution inelastic x-ray scattering and {\itshape ab-initio}
 pseudo-potential calculations.
 Extensive measurements of several samples ($y\sim$0, 0.1 and 0.3) at
 temperatures spanning the magnetic ordering temperature
 ($T_N\sim$145 K for $y\sim$0) and the superconducting transition
 temperature ($T_c$ = 36 K for $y\sim$0.1 and $T_c$ = 45 K for
 $y\sim$0.3) show that there are some changes in phonon spectra with
 temperature and/or doping.
 We compare our measurements with several {\itshape ab initio}
 pseudo-potential models (nonmagnetic tetragonal, oxygen-deficient
 O$_{7/8}$ supercell, magnetic orthorhombic, and magnetic tetragonal)
 and find that the experimentally observed changes are much smaller than
 the differences between the experimental data and the calculations.
 Agreement is improved if magnetism is included in the calculations via
 the local spin density approximation, as the Fe atomic motions parallel to
 the ferromagnetic ordering direction are softened.
 However, the antiferromagnetically polarized modes remain hard, and in
 disagreement with the experimental data.
 In fact, given the increasing evidence for anisotropy in the iron
 pnictide materials, the phonon response is surprisingly isotropic.
 We consider several modifications of the {\itshape ab initio}
 calculations to improve the agreement with the experimental data.
 Improved agreement is found by setting the matrix to zero (clipping the
 bond) between nearest-neighbor antiferromagnetically aligned Fe atoms
 in the magnetic calculation, or by softening only the {\itshape
 in-plane} nearest-neighbor Fe-As force constant in the nonmagnetic
 calculation.
 We discuss these results in the context of other measurements,
 especially of phonons, for several FeAs systems.
 Fluctuating magnetism may be a partial explanation for the failure of
 the calculations, but seems incomplete in the face of the similarity of
 the measured phonon response in all the systems investigated here
 including those known to have static magnetism.
\end{abstract}

\pacs{74.25.Kc, 74.70.Xa, 78.70.Ck}

\maketitle

\section{Introduction}

\begin{figure}
 \includegraphics[keepaspectratio,width=0.3\textwidth]{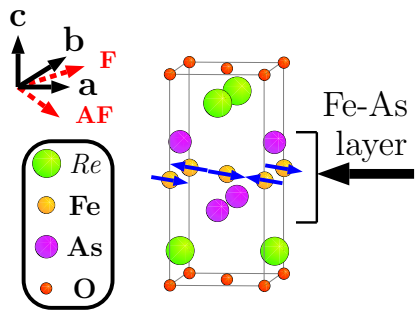}%
 \caption{\label{Fig:unit} (Color online)
 Crystal structure of $Re$FeAsO ($Re$: rare earth).
 The arrows on Fe atoms show the magnetically ordered pattern observed
 for $T<T_N$
 (Refs.~\onlinecite{KamiharaY08a,CruzC08a,KimberSAJ08a,ZhaoJ08a}).
 Dashed (red) arrows indicate the direction of ferromagnetic (F)
 [antiferromagnetic (AF)] alignment of iron spins in the plane.
}
\end{figure}

Since the discovery of superconductivity in LaFeAs(O,F) at 26 K
(Ref.~\onlinecite{KamiharaY08a}), iron-arsenides and related compounds
have been the subject of enormous scientific attention.
These compounds have a layered structure consisting of iron and
pnictide/chalcogenide atoms (``Fe-As layer'') with each iron atom
surrounded by a tetrahedron of pnictide/chalcogenide atoms
(c.f. Fig.~\ref{Fig:unit}).
The iron-based superconductors have been classified into, mainly, four
groups according to the kind of buffer between the Fe-As layers:
$R$FeAs$X$ ($R$: rare earth, $X$ = O or F) with ZrCuSiAs type structure
(``1111'')\cite{KamiharaY08a,RenZA08a,KitoH08a,WuG09a}, $R$Fe$_2$As$_2$
($R$ = Ba, Sr, Ca) with ThCr$_2$Si$_2$ type structure
(``122'')\cite{RotterM08a,SasmalK08a}, $A$FeAs ($A$ = Li, Na) with
Cu$_2$Sb type tetragonal structure
(``111'')\cite{WangXC08a,ParkerDR09a}, and $\alpha$-Fe$Ch$ ($Ch$ = S,
Se, Te) with $\alpha$-PbO type structure
(``11'')\cite{HsuFC08a,MizuguchiY08b}.
Doping electron or hole carriers into the Fe-As layers, or application
of pressure (external or chemical) results in superconductivity at low
temperature.
Some of the ``1111'' type materials show superconductivity
above 50 K (Refs.~\onlinecite{RenZA08a,KitoH08a,WuG09a}), which is the
highest observed superconducting transition temperature ($T_c$) outside
of the copper oxide family of high-$T_c$ superconductors.
Recently more complex compounds having a buffer layer of perovskite
structure have been synthesized pursuing a higher
$T_c$
(Refs.~\cite{OginoH09a,ZhuX09a,KawaguchiN10a,OginoH10b,OginoH10c,ShiragePM10b}).
However, the highest $T_c$ is still achieved in a ``1111'' material.

It has been suggested by many groups that the physical properties of
iron-arsenic compounds are very sensitive to the crystal structure,
especially to the pnictide/chalcogenide height above the Fe plane and/or
the angle of the As-Fe-As
bonds\cite{LeeCH08a,HanMJ09a,KurokiK09a,HoriganeK09a,CruzC10a,NomuraT09a}.
This is further emphasized by the possibility to dramatically affect the
superconducting properties of some material by applying pressure, with
$T_c$ increased in LaFeAs(O,F) from 26 to 43 K under an applied pressure
of 4 GPa (Ref.~\onlinecite{TakahashiH08a}), and in Fe$_{1.01}$Se from
8.5 to 36.7 K under a pressure of 8.9 GPa
(Ref.~\onlinecite{MedvedevS09a}).
This sensitivity suggests the possibility of large coupling between
lattice motion and electronic or, perhaps, magnetic structure.
Meanwhile, early first principles calculations of phonon properties
suggested that the superconductivity in iron-arsenic compounds is not
phonon mediated\cite{SinghDJ08a,BoeriL08a,HauleK08a}.
However, later calculations\cite{MazinII08a,YildirimT09a} demonstrated
the strong dependence of the iron magnetic moment on the
pnictide/chalcogenide atomic position, suggesting some interplay between
phonons and magnetism, while others have noted increased electron-phonon
in magnetic calculations by factors of $\sim$2
(Ref.~\onlinecite{BoeriL10a}) to $\sim$10
(Ref.~\onlinecite{YndurainF11a}).

Experimentally phonons have been measured using various methods.
Inelastic neutron and/or x-ray scattering (INS/IXS) is a powerful tool
to investigate phonon structures, and many measurements have been
already
reported\cite{QiuY08a,ChristiansonAD08a,FukudaT08a,TaconML08a,MittalR08a,MittalR08b,MittalR09c,MittalR09a,PhelanD09a,LeeCH10a,ReznikD08a,MittalR09b,HahnSE09a,ReznikD09a,TaconML09a,TaconML10a,ZbiriM10a}.
Many of these reported small changes with doping and/or temperature.
The experimental phonon data are compared to first principles
calculations, and the introduction of magnetic effect is thought to be
important\cite{ReznikD08a,MittalR09b,HahnSE09a,ReznikD09a,TaconML09a,TaconML10a,ZbiriM10a}.
We will discuss them in detail later.
Here we note that with the exception of Refs.~\onlinecite{FukudaT08a}
and \onlinecite{TaconML09a} all single crystal work has focused on the
``122'' samples, despite their lower $T_c$, because the ``1111''
materials remain relatively difficult to grow.

In the present paper, we report phonon dispersion measurements along
various symmetry directions measured on single crystals of
PrFeAsO$_{1-y}$ using IXS.
This is a ``1111'' compound that shows relatively high $T_c$ up to 45 K
for the samples discussed here and 49 K for the Pr family.
We also perform several first principles calculations, and compare them
carefully with the experimental data.
Our detailed observation confirmed the {\itshape isotropy} of phonon
dispersion even in the antiferromagnetically ordered phase compared to
the first principles calculation considering the magnetic effect.
Though some similar observations, calculations, and the comparisons have
been already discussed for other iron-arsenide
compounds\cite{QiuY08a,ChristiansonAD08a,FukudaT08a,TaconML08a,MittalR08a,MittalR08b,MittalR09c,MittalR09a,PhelanD09a,LeeCH10a,ReznikD08a,MittalR09b,HahnSE09a,ReznikD09a,TaconML09a,TaconML10a,ZbiriM10a},
the similarity of our results to other works ensures that the present
work contains the general features of iron-arsenides.

We then consider several modifications of the first principles
calculations to better interpret our results.
Two modified models are found to have good agreement with the
experimental data.
Interestingly, one is the model that adds additional in-plane
anisotropy, while the other preserves the isotropy of our {\itshape
nonmagnetic} first principles calculation.

This paper is organized as follows.
The details of samples investigated, IXS measurements, and first
principles calculations are provided in Sec.~\ref{Sec:exp_and_calc}.
The experimental results and the comparison with calculations are given
in Sec.~\ref{Sec:results}.
Several modification model of first principles calculations are detailed
in Sec.~\ref{Sec:models}.
Section \ref{Sec:discussion} is dedicated to the discussion of the
results and a summary is given in Sec.~\ref{Sec:conclusion}.

\section{Experiments and Calculations}\label{Sec:exp_and_calc}
\subsection{Samples}

\begin{figure}
 \includegraphics[keepaspectratio,width=0.5\textwidth]{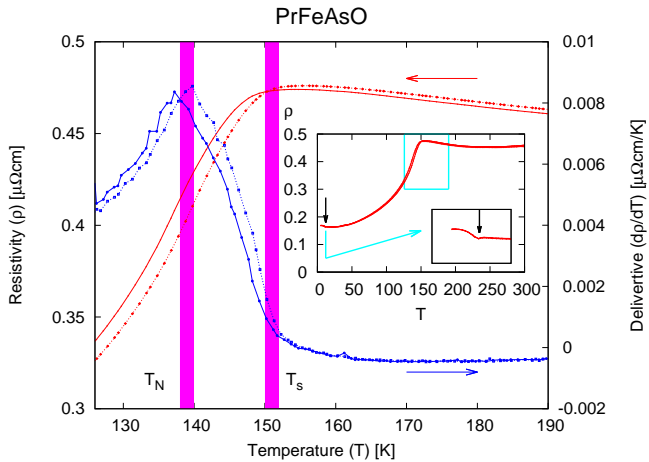}%
 \caption{\label{Fig:registivity} (Color online)
 Electric resistivity ($\rho$) and the derivative ($d\rho/dT$) of a
 PrFeAsO sample near the structural and magnetic transition
 temperatures.
 Solid (dashed) lines correspond to the measurements with decreasing
 (increasing) temperature.
 Inset shows a larger temperature range for $\rho$.
 The deviation in $\rho$ from slow increase with decreasing temperature
 or an abrupt increase in $d\rho/dT$ corresponds to the structural
 transition temperature $T_s$ from tetragonal to orthorhombic symmetry,
 while the inflection point in $\rho$ or the maximum in $d\rho/dT$
 corresponds to the magnetic transition temperature $T_N$.
 A small upturn below 12 K indicated by an arrow in the inset
 corresponds to the ordering of Pr magnetic moments.
}
\end{figure}

The investigated PrFeAsO$_{1-y}$ single crystals are summarized in Table
\ref{Tbl:sample}.
\begin{table}
 \caption{\label{Tbl:sample}PrFeAsO$_{1-y}$ samples.
 The thickness of the samples was a few tens $\mu$m.
 The sample 'doped-1' was also used in a previous
 study (Ref.~\onlinecite{FukudaT08a}).
 }
 \begin{ruledtabular}
  \begin{tabular}{cccccc}
   name & $y$       & $T_s$ [K] & $T_N$ [K] & $T_c$ [K] & typical size [$\mu$m] \\ \hline
   parent  & 0.0       & $\sim$149 & $\sim$139 & ---       & $\sim$500 \\
   doped-1 & $\sim$0.1 & ---       & ---       & 36        & $\sim$100 \\
   doped-2 & $\sim$0.3 & ---       & ---       & 45        & $\sim$300
  \end{tabular}
 \end{ruledtabular}
\end{table}
They were prepared using high-pressure growth as described in
Ref.~\onlinecite{IshikadoM09b}.
The first samples were
relatively small $\sim 0.1 \times 0.15 \times 0.02$ mm$^3$, but with
improvements in growth techniques, they became comfortably large,
$\sim$0.5 mm in the $ab$ plane.
Sample thickness varied from about 0.02 to 0.05 mm.
X-ray diffraction on a four-circle diffractometer was used to verify
that all samples were single grains, with, typically, a mosaic spread of
about 1$^\circ$.
The $T_c$'s of the studied superconducting samples were found to be 36
and 45 K by measuring magnetic susceptibility using a superconducting
quantum interference device (SQUID) magnetometer.
The electrical resistivity in the parent sample showed an abrupt change
at $\sim$149 K and the derivative had a maximum at $\sim$139 K
(Fig.\ref{Fig:registivity}).
According to the previous transport measurements in iron
pnictides\cite{HessC09a,RotterM08b}, these temperatures
correspond to the structural and magnetic transition, respectively.
These temperatures are slightly higher than the values given in
Refs.~\onlinecite{KimberSAJ08a} and \onlinecite{ZhaoJ08b}, possibly
because we are more nearly at the precise stoichiometric composition.
There is a small anomaly in $\rho$ around 12 K where the Pr spins
order.
This is nearly the same as previously
reported\cite{ZhaoJ08b,KimberSAJ08a}.

We use tetragonal notation for all samples, with axes along the
next-nearest-neighbor iron atoms.
In the parent sample the lattice becomes orthorhombic below $T_s$,
followed by ordering of the Fe magnetic moments below $T_N$.
When required, we take the ferromagnetically (antiferromagnetically)
ordered direction in the Fe plane to be parallel to $<$110$>$
($<$1\={1}0$>$) as shown in Fig.~\ref{Fig:unit}.
The lattice constant along $<$110$>$ or the ferromagnetically ordered
direction is shorter than that along $<$1\={1}0$>$ or the
antiferromagnetically ordered direction, $a_{<110>}<a_{<1\bar{1}0>}$.

\subsection{IXS measurements and data analysis}

IXS measurements were performed at BL35XU
(Ref.~\onlinecite{BaronAQR00a}) of SPring-8.
The x-ray beam from the storage ring was first monochromatized to
$\sim$eV by a double crystal, liquid nitrogen cooled, Si(1~1~1)
monochromator, and then down to 0.8 meV using a Si(11~11~11)
backscattering monochromator operating at 21.75 keV.
The incident beam was focused, by a bent cylindrical mirror, to a spot
size of about 70 microns in diameter [full width at half maximum (FWHM)]
at the sample position.
The scattered radiation was analyzed using 12 spherical crystals in a 
4(horizontal) $\times$3(vertical) array on a horizontal 10 m 2$\theta$
arm.
Each analyzer focused the radiation into a separate detector located
near the sample.

The samples were mounted on thin glass rods, which were set on a small
goniometer head in air for room temperature measurements, or on the cold
finger of a $^4$He closed-cycle refrigerator for low temperature
investigations.
The penetration depth of PrFeAsO$_{1-y}$ for 21.75 keV x rays is about
60 $\mu$m, which, with our thin samples, enabled us to measure in a
transmission (Laue) geometry for many $Q$ points.
The surface normal of all samples was along the $c$ axis, and they were
aligned with the $<$010$>$ or $<$1\={1}0$>$ directions approximately
vertical.

The use of the two-dimensional analyzer array allowed parallelization of
measurements for either nearly pure longitudinal or transverse modes
dispersing out from the $\Gamma$ point\cite{BaronAQR08a}.
For example, with $<$010$>$ vertical and the $<$100$>$ in the horizontal
scattering plane, we can measure simultaneously four
$Q$ points along approximately $<$100$>$ longitudinal direction or
three $Q$ points along approximately $<$010$>$ transverse direction
from a ($H$~0~0) Bragg points using, respectively, a horizontal or
vertical line of analyzers.
Likewise, $<$110$>$ longitudinal and transverse modes can be
investigated easily with the condition $<$1\={1}0$>$ vertical.
The analyzers were not always perfectly centered along a symmetry
direction, but the deviation was typically very small [$<$0.02
reciprocal lattice units (r.l.u.) in $H$ and $K$, and $<$0.15 r.l.u. in
$L$ at the momentum transfers we measured].
A typical value for the momentum resolution (corresponding to the
analyzer acceptance) is 0.06 r.l.u. in $H$ and $K$, and 0.12 r.l.u. in
$L$.

To analyze the data, the obtained spectra were fit to the sum of a
resolution-limited elastic peak and several Lorentzian phonon peaks:
%
\begin{eqnarray}
 &&I(E)=A\ {\rm Res}(E) \nonumber \\
 &&+\sum_i\frac{B_i\ b_i}{\pi}\left\{\frac{n(E+E_i)}{(E+E_i)^2+b_i^2}+\frac{n(E-E_i)+1}{(E-E_i)^2+b_i^2}\right\},
  \label{Eq:fit_fn}
\end{eqnarray}
%
where Res$(E)$ is the instrumental resolution function measured using
elastic scattering from plexiglas, $A$ and $B_i$ are constants, $E_i$
and $b_i$ are the energy and the width of the $i$-th phonon peak, and
$n(E)$ is the Bose thermal factor.
Fitting to the above function was done using a nonlinear
Levenberg-Marquardt least squares algorithm.

\subsection{First principles calculations}

Several first principles calculations of PrFeAsO$_{1-y}$ and LaFeAsO
were carried out using {\textsc
VASP}\cite{KresseG93a,KresseG96b,KresseG96a} and a projector
augmented-wave (PAW) method\cite{BlochlPE94a,KresseG99a}, in the
generalized gradient approximation (GGA), and local spin density
approximation (LSDA). (Table~\ref{Tbl:calc})
\begin{table*}
 \caption{\label{Tbl:calc}
 Structural parameters from first principles calculations (upper
 portion) and the experiment (lower portion).
 ``original'' is the simplest calculation with nonmagnetic tetragonal
 structure, while ``O$_{7/8}$'' is a supercell calculation with a 12.5\%
 ordered oxygen deficiency.
 ``magnetic'' and ``mag. tetra'' are the first principles calculations
 with antiferromagnetically ordered Fe moments (after
 Ref.~\onlinecite{CruzC08a}), and the crystal symmetry is orthorhombic
 and tetragonal ($a_{ortho}=b_{ortho}$; {\itshape Ibam}), respectively.
 ``O$_{7/8}$'' gives different $z_{Pr}$'s and $z_{As}$'s and the
 averages are given.
 In the parent material below $T_N$ Fe spin moments are aligned
 ferromagnetically along the $b_{ortho}$ direction
 ($\parallel <$110$>$, the notation here as shown in Fig.~\ref{Fig:unit}).
 }
 \begin{ruledtabular}
  \begin{tabular}{llllllllll}
   name       & compound         & symmetry                    &
      $a$ [{\AA}] & $a_{ortho}$/$b_{ortho}$ [{\AA}] & $c$ [{\AA}] &
      $z_{Pr/La}$   & $z_{As}$      & $T$ [K] & ~ \\ \hline
   original   & PrFeAsO          & {\itshape P4/nmm} (tetra)   &
      4.0124      & (5.6744)                        & 8.4863      &
      0.14472       & 0.64052       & (0)     & ~ \\
   O$_{7/8}$  & PrFeAsO$_{7/8}$  & {\itshape P\={4}m2} (tetra) &
      3.98635       & 5.6375                          & 8.4336      &
      $<$0.15466$>$ & $<$0.64367$>$ & (0)     & ~ \\
   magnetic   & LaFeAsO          & {\itshape Ibam} (ortho)     &
      ~           & 5.7320/5.6616                   & 8.6437      &
      0.14359       & 0.64666       & (0)     & ~ \\
   mag. tetra & LaFeAsO          & {\itshape Ibam} (tetra)     &
      (4.0301)      & 5.69937                         & 8.7368      &
      0.14269       & 0.64636       & (0)     & ~ \\ \hline
   parent     & PrFeAsO          & ~                           &
      3.976       & (5.623)                         & 8.572       &
      ~             & ~             & R.T. & ~ \\
   doped-1    & PrFeAsO$_{0.9}$  & ~                           &
      3.976(1)    & (5.623)                         & 8.5686(2)   &
      ~             & ~             & R.T. & ~ \\
   doped-2    & PrFeAsO$_{0.7}$  & ~                           &
      3.961       & (5.601)                         & 8.539       &
      ~             & ~             & R.T. & ~ \\
   Powder & PrFeAsO          & {\itshape P4/nmm} (tetra)       &
      3.97716(5)  & (5.6246)                        & 8.6057(2)   &
      0.1397(6)     & 0.6559(4)     & 175  & Ref.~\onlinecite{ZhaoJ08a} \\
   ~          & PrFeAsO          & {\itshape Cmma} (ortho)     &
      ~           & 5.6374(1)/5.6063(1)             & 8.5966(2)   &
      0.1385(5)     & 0.6565(3)     & 5    & Ref.~\onlinecite{ZhaoJ08a} \\
   ~          & PrFeAsO$_{0.85}$ & {\itshape P4/nmm} (tetra)   &
      3.9686(1)   & (5.6124)                        & 8.5365(3)   &
      0.1450(7)     & 0.6546(5)     & 5    & Ref.~\onlinecite{ZhaoJ08a} \\
  \end{tabular}
 \end{ruledtabular}
\end{table*}
Phonons were calculated via the {\textsc PHONON} package using a direct
method\cite{ParlinskiK97a}.
The supercell size used for each calculation is $2a \times 2a \times c$
(32 atoms) for ``original'', $2a \times 2a \times c$ (31 atoms) for
``O$_{7/8}$'', and $2\sqrt{2}a \times 2\sqrt{2}a \times 2c$ (128 atoms)
for both ``magnetic'' and ``mag. tetra,'' where $a$ and $c$ are lattice
constants of the tetragonal primitive unit cell
(cf. Table~\ref{Tbl:calc}).
Total energies and inter-atomic forces were calculated for 16, 56, 30,
and 30 symmetry-inequivalent displacements for original, O$_{7/8}$,
magnetic and mag. tetra calculations, respectively.
The energy cut-off for plane waves was 550 eV, the spacing of $k$ points
was less than 0.1 {\AA}$^{-1}$ and the convergence condition was that
the total energy difference be less than 1 $\mu$eV.
All structures were relaxed.
The structural parameters are summarized in Table~\ref{Tbl:calc}.
Stripe antiferromagnetic ordering, as observed experimentally, is
stabilized in the magnetic calculations, with structural parameters that
are nearer to the experimentally determined values\cite{IshibashiS08a}.

Magnetic calculations were done using the La pseudopotential
instead of Pr to avoid difficulties in treating the spins of the
localized Pr $f$ electrons in the core pseudopotential.
However, a comparison of the results of nonmagnetic calculations using
La and Pr showed the results to be nearly identical for the phonons,
with the differences between the two calculations being much smaller
than those between the calculation and the measured data.
Considering a purely mass difference would suggest phonon frequency
changes of less than 1 percent for La/Pr modes.

\section{Results}\label{Sec:results}
\subsection{IXS measured data}\label{Sec:IXS_meas}

\begin{figure}
 \includegraphics[keepaspectratio,width=0.5\textwidth]{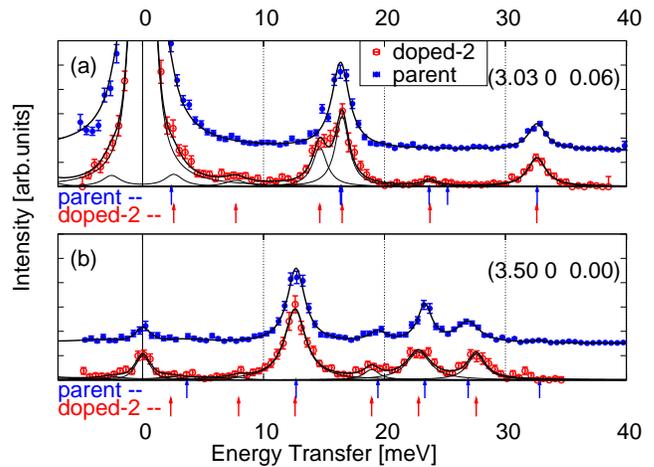}%
 \caption{\label{Fig:spect_typ} (Color online)
 Typical IXS spectra of doped PrFeAsO$_{1-y}$ (doped-2; open circle) and
 parent PrFeAsO (closed circle; shifted vertically for clarify) at room
 temperature (a) near $\Gamma$ point ($Q$ = (3.03~0~0.06)) and (b) at
 Brillouin zone boundary ($Q$ = (3.50~0~0.00)).
 The lines show the fit to the data using Eqn.~(\ref{Eq:fit_fn}).
 The individual peaks are also shown for the doped-2.
 The arrows below each plot show the peak positions based on the fits.
 Generally the elastic intensity is larger near to the $\Gamma$ point.
}
\end{figure}

Figure~\ref{Fig:spect_typ} shows typical IXS spectra of parent PrFeAsO
and doped PrFeAsO$_{1-y}$ (doped-2; $T_c$=45 K) at room temperature at
$Q$ = (3.03~0~0.06) (near the Brillouin zone center, $\Gamma$ point) and
(3.50~0~0.00) (Brillouin zone boundary).
An elastic peak and several phonons were observed, and fit 
using the function of Eq.~(\ref{Eq:fit_fn}).
The sum fit curve as well as individual phonon lineshapes are shown in
Fig.~\ref{Fig:spect_typ}.

\begin{figure}
 \includegraphics[keepaspectratio,width=0.5\textwidth]{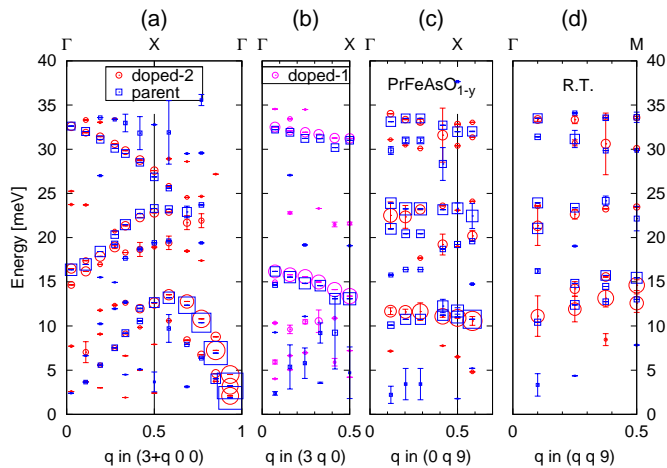}%
 \caption{\label{Fig:dis_exp_dope} (Color online)
 Dispersion relations along some high-symmetry directions of
 doped PrFeAsO$_{1-y}$ (open circle) and parent PrFeAsO (closed circle)
 at room temperature.
 The $Q$ = (3~$q$~0) data of (b) are from the sample with $y\sim$0.1
 ($T_c=$ 36 K; doped-1), and the other ones are from the sample with
 $y\sim$0.3 ($T_c=$ 45 K; doped-2).
 The area of the symbols is proportional to the peak intensity [$B_i$ in
 Eq.~(\ref{Eq:fit_fn})] and the error bars are
 the intrinsic peak widths estimated by subtracting the experimental
 resolution from the fit result.
}
\end{figure}

Figure~\ref{Fig:dis_exp_dope} shows the dispersion relations of
PrFeAsO$_{1-y}$ (parent: $y=$0, doped-1: $y\sim$0.1 or doped-2:
$y\sim$0.3)
at room temperature along some high-symmetry directions;
$Q$ = (3+$q$~0~0), (3~$q$~0), (0~$q$~9), and ($q$~$q$~9).
The area of the symbols is proportional to the integrated peak
intensity, $B_i$, and the error bars give the intrinsic width, after
subtracting the measured instrumental resolution width from $|b_i/2|$.
The uncertainty in mode energy is $<$0.5 meV except for some small
and/or broad peaks.
Though the data in doped PrFeAsO$_{1-y}$ having slightly low $T_c$
(doped-1) are plotted only in Fig.~\ref{Fig:dis_exp_dope}(b), the
doped-1 and doped-2 samples show exactly the same result at
$Q$ = (3+$q$~0~0) (not shown).

There is a doping dependence, which is also branch dependent.
For example, there are three strong branches observed at $Q$ =
(3+$q$~0~0) in Fig.~\ref{Fig:dis_exp_dope} (a).
With doping, the branch dispersing from $\sim$32 meV at
$Q$ = (3~0~0) to $\sim$27 meV at (3.5~0~0) hardens, while the branch
dispersing from $\sim$16 meV at (3~0~0) to $\sim$23 meV at (3.5~0~0)
softens slightly.
The hardening/softening may be larger at the zone boundary than at the
zone center, which can be seen more clearly in
Fig.~\ref{Fig:spect_typ}.
However, the differences are small ($\sim$0.5 meV), and there is no
drastic change in the overall dispersion between the parent and doped
sample data.
The doping dependence in dispersion relations at low temperature below
$T_N$ and $T_c$ show features similar to Fig.~\ref{Fig:dis_exp_dope}
(not shown).

\begin{figure}
 \includegraphics[keepaspectratio,width=0.5\textwidth]{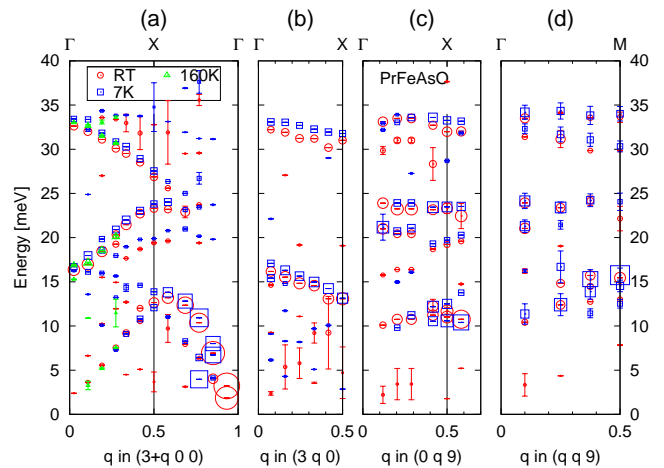}%
 \caption{\label{Fig:dis_exp_T} (Color online)
 Dispersion relations along some symmetry directions for the
 parent PrFeAsO.
 The figure style is the same as that of Fig.~\ref{Fig:dis_exp_dope}.
 The data at room temperature are identical to those in
 Fig.~\ref{Fig:dis_exp_dope}.
}
\end{figure}

Figure~\ref{Fig:dis_exp_T} shows the temperature dependence of the
dispersion in the parent PrFeAsO at $Q$ = (3+$q$~0~0), (3~$q$~0),
(0~$q$~9), and ($q$~$q$~9).
We can see small changes in some high energy branches.
The mode energy increases with
decreasing temperature, as may just be the result of  thermal contraction.
Measurements about $T_s$ (and $T_N$) near $Q$ = (3~0~0)
[Fig.~\ref{Fig:dis_exp_T}(a)], did not show any strong changes on
crossing the transition temperature, with mode frequencies being nearly
unchanged and most mode intensities following the usual Bose factor.
That is, the temperature dependence in doped superconducting
PrFeAsO$_{1-y}$ is similar to that in the parent sample, with a slight
hardening as temperature is decreased, and no drastic change in the
overall dispersion.

\begin{figure}
 \includegraphics[keepaspectratio,width=0.5\textwidth]{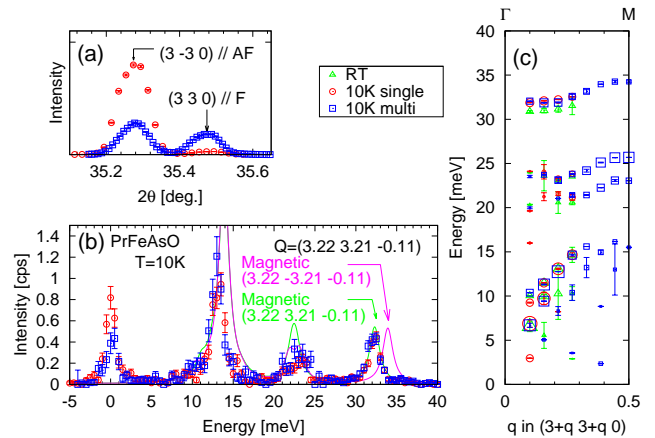}%
 \caption{\label{Fig:dis_exp_110} (Color online) No magnetic splitting
 is observed in the phonon spectra for PrFeAsO at 10 K.
 (a) 2$\theta$ scans near (3~3~0)/(3~\={3}~0) Bragg point.
 The (3~3~0)/(3~\={3}~0) correspond to the scattering from
 antiferromagnetically/ferromagnetically ordered (orthorhombic $a$/$b$)
 direction, respectively.
 (b) Phonon spectra at $Q$ = (3.22~3.21~--0.11).
 The lines are the spin polarized {\itshape ab initio} ``magnetic''
 orthorhombic calculation, (see Sec.~\ref{Sec:abinitio}) convoluted with
 the instrumental resolution.
 (c) Dispersion relations.
}
\end{figure}

Measurements of the parent PrFeAsO were made along the $<$110$>$
direction at 10 K, well below the measured $T_N\simeq$ 139 K, to
investigate possible effects of the orthorhombic structural distortion
and appearance of magnetism.
This direction potentially twins, with the $<$110$>$ and $<$1\={1}0$>$
directions becoming distinct.
If the beam hits a twinned section of the sample, it would be reasonable
(especially based on the magnetic calculations discussed below) to
expect phonon splitting as the modes polarized in the $<$110$>$
ferromagnetic ordering direction (shorter lattice constant) could have
different energies from those in the $<$1\={1}0$>$ antiferromagnetic
ordering direction.
With some effort, we were able to find part of the sample where twinning
was clearly observed, based on the split of the (3~3~0) Bragg reflection
[Fig.~\ref{Fig:dis_exp_110}(a)], however, even in this twinned region,
the phonon splitting near $\sim$34 meV expected to be $\sim$3 meV from
magnetic calculation did not appear as shown in
Fig.~\ref{Fig:dis_exp_110}(b).
The dispersion is plotted in Fig.~\ref{Fig:dis_exp_110}(c) for the
parent sample from two regions at low temperature as well as those at
room temperature.
The spectral shape, and the dispersion is essentially identical.
One notes that there is some hardening with decreasing temperature
observed in the high energy branch [Fig.~\ref{Fig:dis_exp_110}(c)] and
almost no doping dependence in this direction (not shown).
This small doping/temperature effects are similar to other
phonon modes discussed above (Figs.~\ref{Fig:dis_exp_dope} and
\ref{Fig:dis_exp_T}).

\subsection{Comparison with {\itshape ab initio} calculations}
\label{Sec:abinitio}

\begin{figure}
 \includegraphics[keepaspectratio,width=0.5\textwidth]{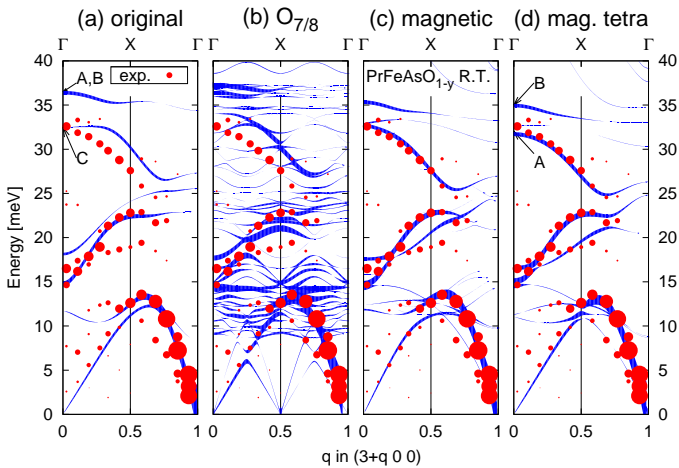}%
 \caption{\label{Fig:dis_H00} (Color online)
 Comparison of the measured dispersion for PrFeAsO$_{1-y}$ ($T_c=$ 45 K;
 doped-2) at room temperature at $Q$ = (3+$q$~0~0) and
 various first principles calculations.
 The same data are repeated in each panel for comparison with the
 different calculations:
 (a) ``original'' calculation using a nonmagnetic tetragonal structure,
 (b) the ``O$_{7/8}$'' calculation with an oxygen deficiency,
 (c) the ``magnetic'' first principles calculation,
 and (d) the tetragonal magnetic calculation.
 See also the text and table \ref{Tbl:calc} for details.
 The area of the experimental data points (line height of the
 calculations) show the measured (calculated) mode intensity for IXS
 spectra.
 For clarity, a cutoff has been used in plotting acoustic mode intensity
 which becomes large near Bragg points.
 The low energy acoustic mode appearing at ($4-q$~0~0) is probably a
 transverse mode.
 See text at the end of Sec.~\ref{Sec:abinitio}.
}
\end{figure}

We now compare the dispersion against the unmodified {\itshape ab
initio} calculations.
Since the effect of carrier doping, or sample temperature, on the
phonon spectra is relatively small (see Figs.~\ref{Fig:dis_exp_dope} and
\ref{Fig:dis_exp_T}) we plot only one data set, that from the
superconducting PrFeAsO$_{1-y}$ (doped-2).
Figure~\ref{Fig:dis_H00} shows the comparison between the experimental
data and the calculations at $Q$ = (3+$q$~0~0).
Again, the area of experimental data points shows the mode intensity
after removal of the Bose factor.
For the calculations, the line height shows the expected intensity.
The scale factor (symbol area to line height) is fixed for all data
sets.
The experimental data in Fig.~\ref{Fig:dis_H00} are the same as shown
previously [doped-2 in Fig.~\ref{Fig:dis_exp_dope}(a)], but the error
bars indicating the intrinsic peak width are omitted.

The agreement between the data and calculation is fairly good for low
energy branches.
However, the branch dispersing from $\sim$33 meV at the $\Gamma$ point
shows notable differences between the different calculations.
The measured energy is generally significantly lower than that of the
first principles calculation with a nonmagnetic ground state
[Fig.~\ref{Fig:dis_H00}(a) ``original''], consistent with our earlier
work\cite{FukudaT08a}.
As the superconducting samples are oxygen deficient, we also made
calculations with a deficiency (12.5 \%) using a large unit cell size
2 $\times$ 2 $\times$ 1 with one oxygen removed
[Fig.~\ref{Fig:dis_H00}(b) ``O$_{7/8}$''].
The agreement with the experimental data is similar to the ``original''
calculation, but many added branches appear, as might be expected from
adding an ordered deficiency.
Such additional modes are not observed in our measurement, and the
agreement with the high energy branch in the ``original'' calculation is
not improved.
This calculation then mostly provides confirmation that the oxygen
deficiency probably does not have a surprising impact on the detailed
phonon dispersion.
This is consistent with the recent first principles calculation on the
doping dependence of phonon DOS in LaFeAsO$_{1-x}$F$_x$ using the
virtual crystal approximation\cite{YndurainF11a}.

We also carried out calculations with antiferromagnetically ordered
Fe moments in an orthorhombic structure, essentially that of de la Cruz
{\itshape et al.}\cite{CruzC08a}.
The magnetic order of Pr atoms was not considered as they were
substituted with La in the calculation.
The result is plotted in Fig.~\ref{Fig:dis_H00}(c) (magnetic).
The agreement with the experimental data is better than the
``original'' model.
We also performed a calculation of a tetragonal magnetically ordered
material to distinguish between the effects of the magnetic moment and
crystal symmetry.
The result is plotted in Fig.~\ref{Fig:dis_H00}(d) (mag. tetra), and
it is very similar to the ``magnetic'' calculation.
This clearly shows the phonon softening of the branch at 27--33 meV in
the first principles calculation comes from allowing the magnetic order
of the iron atoms: the orthorhombic/tetragonal crystal symmetry change
has a relatively small effect on the calculated phonon dispersion.

\begin{figure}
 \includegraphics[keepaspectratio,width=0.4\textwidth]{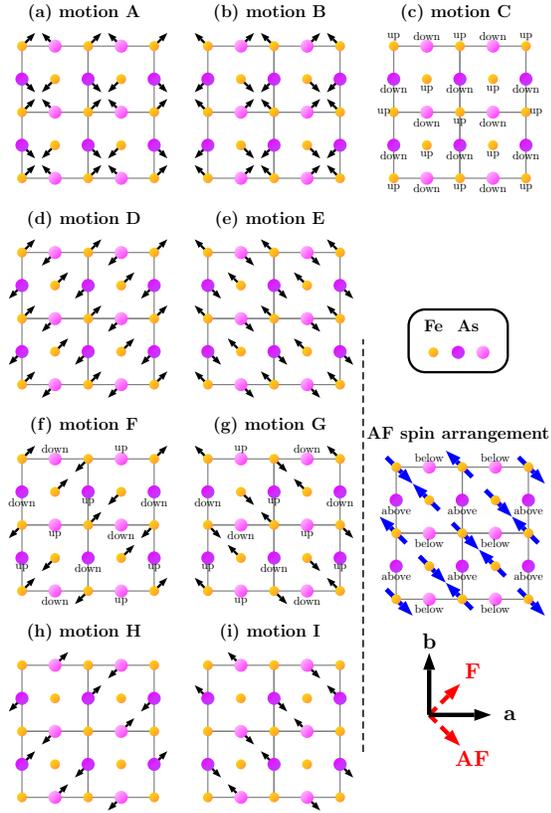}%
 \caption{\label{Fig:motion} (Color online)
 The bottom right is a schematic of the Fe-As layer:
 As atoms are located above or below the Fe atomic plane as specified.
 The arrows on Fe atoms specify the magnetic spin directions for
 ``magnetic'' and ``mag. tetra'' calculations, where the ferromagnetic
 (antiferromagnetic) propagation vector corresponds to be $<$110$>$
 ($<$1\={1}0$>$) as shown in the red dashed arrow F (AF).
 In (a)--(i) arrows specify atomic displacement patterns in an Fe-As
 layer at several points in Figs.~\ref{Fig:dis_H00} and
 \ref{Fig:dis_HK0_330}.
 Atomic movements along out-of-plane direction are specified by ``up''
 or ``down'' near the atoms.
}
\end{figure}

The addition of the magnetism, at first glance, improves the agreement
between the data and the calculations.
It is worth emphasizing that this is true for both the parent material
(data not shown) which explicitly shows magnetic order, and also the
superconducting materials [data as shown in Fig.~\ref{Fig:dis_H00}(c)]
which do not show evidence of static magnetic order.
The latter is somewhat surprising, and we will discuss it later again.
However, while the magnetic calculations do better, they also predict
splitting of modes that is much larger than that observed in our data.
This is evident in Figs.~\ref{Fig:dis_H00}(c) and (d), where
calculations give a high energy branch ($\sim$34 meV) that is not
observed here.
The high energy branch originates from the magnetic calculation lifting
a degeneracy between the ferromagnetically and antiferromagnetically
polarized modes:
the measured phonon energies are consistent with the energies calculated
for modes with Fe motions in the ferromagnetic ordering direction
[motion A in Fig.~\ref{Fig:motion}(a)] but not with the calculated
energies of modes with Fe motions in the antiferromagnetic direction,
of which should also appear in the ($3+q$~0~0) direction [motion B in
Fig.~\ref{Fig:motion}(b)].

The failure of the magnetic calculations is further confirmed by
investigating dispersion along the $<$110$>$ direction.
A splitting similar to that mentioned above is calculated to appear
between ferromagnetically polarized modes (observed in a longitudinal
$<$110$>$ geometry) and antiferromagnetically polarized modes (observed
in the $<$1\={1}0$>$) in a twinned portion of the parent below $T_N$.
As discussed previously, Fig.~\ref{Fig:dis_exp_110} shows the
measurement of longitudinal modes in the $<$110$>$ direction, from both
a single domain and a twinned section of the crystal.
The phonon spectra are essentially identical, showing no evidence of the
calculated splitting.

\begin{figure}
 \includegraphics[keepaspectratio,width=0.5\textwidth]{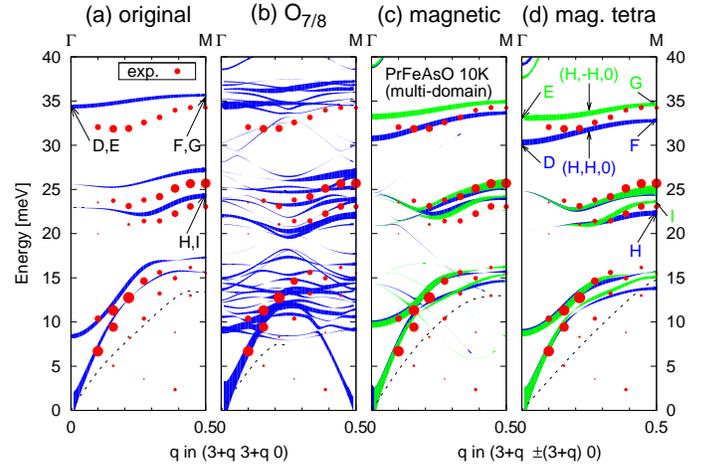}%
 \caption{\label{Fig:dis_HK0_330} (Color online)
 Dispersion along $Q$ = ($3+q$~$3+q$~0) for the parent sample PrFeAsO.
 In this direction of (c) ``magnetic'' and (d) ``mag. tetra''
 calculations there are two domains originated from crystal structure
 and/or magnetic order, and the phonon dispersions from each domains are
 plotted.
 See the caption of Fig.~\ref{Fig:dis_H00} for details.
 The weak, low energy, acoustic mode that can be seen in the data is a
 transverse acoustic mode (dashed line) that appears in the measured
 spectra, probably due to the finite momentum acceptance of the
 analyzers.
 See the text at the end of Sec.~\ref{Sec:abinitio} for discussion.
}
\end{figure}

The lack of splitting is also confirmed in Fig.~\ref{Fig:dis_HK0_330}
along ($3+q$~$3+q$~0) corresponding to Fig.~\ref{Fig:dis_exp_110}(c).
In the figure, the experimental data are compared with (a) ``original'',
(b) ``O$_{7/8}$'', (c) ``magnetic'', and (d) ``mag. tetra''
calculations.
The $<$110$>$ and $<$1\={1}0$>$ directions are not
equivalent, as is shown by choosing different colors (gray scales).
The general features are consistent with those of $Q$ = (3+$q$~0~0) in
Fig.~\ref{Fig:dis_H00}, in that the branches at low energy are explained
fairly well by any calculation, while the branches at the energy of
30-35 meV are not.
A similar feature is also obtained in the data along other symmetry
directions of $Q$ = (3~$q$~0), (0~$q$~9) and ($q$~$q$~9), which are
corresponding to (b)--(d) in Figs.~\ref{Fig:dis_exp_dope} and
\ref{Fig:dis_exp_T}.

Figure~\ref{Fig:motion} investigates the calculated polarizations
related to the splitting of ferromagnetically and antiferromagnetically
polarized modes in more detail, showing the atomic motions at $\Gamma$
and M points.
For these modes only Fe and As atoms move, while rare-earth (La/Pr) and
O atoms do not.
The motions A and B are degenerate in the ``original'' calculation, and
split in ``magnetic'' and ``mag. tetra'' calculations.
The pairs D and E, F and G, and H and I are the same as the pair A and
B.
In each pair the phonon energy of motions A, D, F, and H is lower
than that of motions B, E, G, and I.
In all cases, the motion containing the vibration of Fe atoms along the
$<$110$>$ direction (ferromagnetic mode) has lower energy than that
along the $<$1\={1}0$>$ direction (antiferromagnetic mode) excepting
those modes without in-plane Fe motion (i.e., H and I where Fe atoms do
not vibrate).

Finally we note the appearance of a transverse mode in the longitudinal
spectra, as can be seen in Fig. \ref{Fig:dis_HK0_330}.
The predominantly transverse character of the mode is confirmed by its
higher intensity in the "off symmetry" analyzers (data not
shown) which have a larger transverse contribution.
Its appearance, in principle, then can be explained by the finite
momentum acceptance of the analyzers [for example $\Delta Q\simeq$ 0.029
r.l.u. near the $Q$ = (3~0~0)].
However, the investigation of the calculations in both this section, and
the next, shows that there is also the potential that the mode itself
may be of mixed polarization.
Furthermore, the amount of mixing
might be some way of selecting between different calculations, though
such a detailed analysis, relying heavily on both the intensity
and momentum resolution is beyond the scope of the present work.

\section{Comparison with modified models}\label{Sec:models}

\begin{figure}
 \includegraphics[keepaspectratio,width=0.4\textwidth]{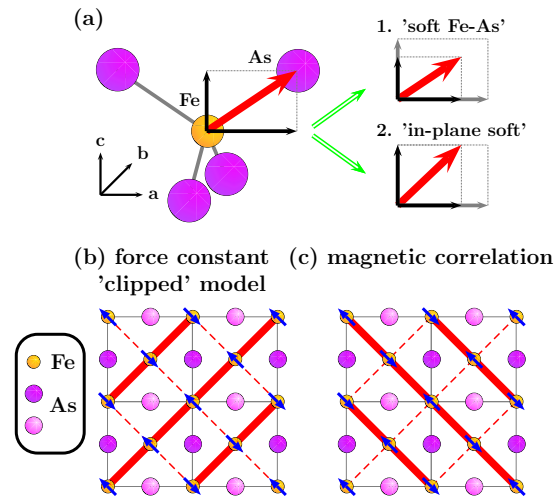}%
 \caption{\label{Fig:motion2} (Color online)
 Schematic of model modifications.
 (a) The ``soft Fe-As'' model softens both the in-plane and the
 out-of-plane components, while the ``in-plane soft'' reduces only the
 in-plane correlation.
 (b) Strength of the force constants between Fe atoms on an Fe-As layer
 in the ``clipped'' model.
 A strong correlation exists along the ferromagnetic direction as shown
 by the thick lines.
 (c) Effective magnetic exchange coupling constants between Fe atoms in
 an Fe-As layer suggested by spin wave measurement in
 SrFe$_2$As$_2$ (Ref.~\onlinecite{ZhaoJ08c}) and CaFe$_2$As$_2
 (Ref.~$\onlinecite{ZhaoJ09a}).
 Strong interaction (large $J$) is along the direction of
 antiferromagnetic order.
}
\end{figure}

The ``magnetic'' calculations provide the best over-all agreement with
our data out of those presented above.
However, there remain significant discrepancies with the calculations
predicting mode splitting that is not observed experimentally as
discussed above.
To better understand these continued discrepancies, we consider several
direct modifications to the real-space force constant matrices resulting
from (the interpolation of) the {\itshape ab initio} calculations.
Since the discrepancies between the calculations and the data remain
generally larger than those between different dopings or temperatures,
we focus on modifying the models to get globally similar characteristics
to the data.
The models are summarized in Table~\ref{Tbl:model} and
Fig.~\ref{Fig:motion2}.
\begin{table}
 \caption{\label{Tbl:model} Modified model calculations.
 See text for discussion and Fig.~\ref{Fig:motion2}.
 }
 \begin{ruledtabular}
  \begin{tabular}{ccl}
   name & based {\it ab initio} calc. & modification \\ \hline
   soft Fe-As    & ``original'' & weaken Fe-As \\
   in-plane soft & ``original'' & weaken in-plane Fe-As \\
   clipped       & ``magnetic'' & cut Fe-Fe along AF
  \end{tabular}
 \end{ruledtabular}
\end{table}
For all of them we preserved the crystal symmetry, and recalculated the
self forces to keep the translational invariance.
Optimizations were done by hand.

\begin{figure}
 \includegraphics[keepaspectratio,width=0.5\textwidth]{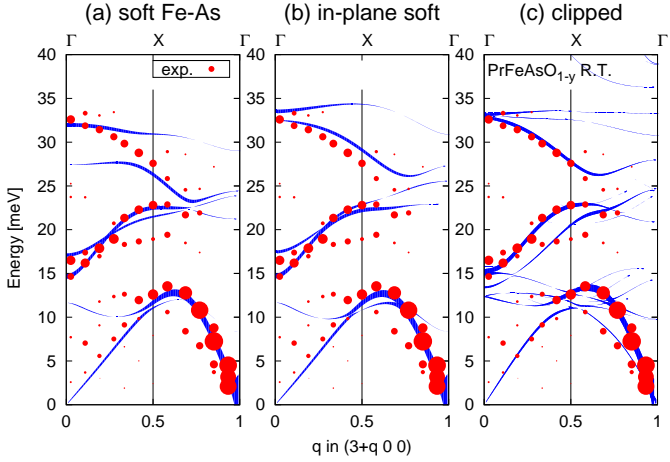}%
 \caption{\label{Fig:dis_comp_model} (Color online)
 (a) Comparison of the experimental data of PrFeAsO$_{1-y}$
 ($T_c$ = 45 K; doped-2) at $Q$ = ($3+q$~0~0) at room temperature with
 the ``soft Fe-As'' model, (b) the comparison with ``in-plane soft''
 model, and (c) the comparison with ``clipped'' model.
 The area of the experimental data as well as the line width of the
 calculations show the peak intensity on IXS spectra.
 See text for details.
}
\end{figure}

The simplest model we consider is the ``soft Fe-As'' model of
Ref.~\onlinecite{FukudaT08a} where the nearest neighbor Fe-As force
constant was scaled (reduced) by 30 \% [Fig.~\ref{Fig:motion2}(a)-1]
from that of the nonmagnetic 'original' calculation.
This was shown to reproduce the softening of the peak in the DOS, but,
not completely the dispersion in our previous work\cite{FukudaT08a}.
The comparison with the in-plane longitudinal mode along $<$100$>$ is
shown in Fig.~\ref{Fig:dis_comp_model}(a).
While the high energy branch is softened, the other branches are almost
unchanged.
The agreement with the experiment improves compared to the ``original''
calculation, but the branch shape is clearly different, with an
anti-crossing in the calculation that is not present in the
measurements.

\begin{figure}
 \includegraphics[keepaspectratio,width=0.5\textwidth]{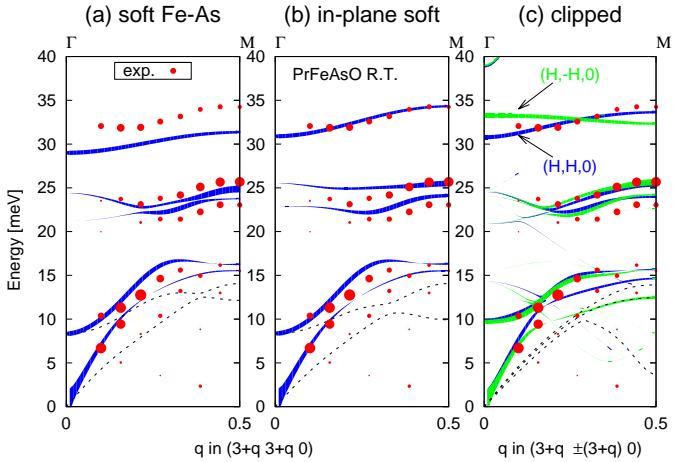}%
 \caption{\label{Fig:dis_comp_model_HK0_330} (Color online)
 As Fig.~\ref{Fig:dis_comp_model} but the data are from parent
 PrFeAsO and along $Q$ = ($3+q$~$3+q$~0).
 In (c) ``clipped'' model the $<$110$>$ and $<$1\={1}0$>$ directions are
 inequivalent and are shown in different colors (gray scales).
 }
\end{figure}

It is possible to modify the nonmagnetic tetragonal (``original'')
calculation to agree better with the observed dispersion by softening
only the in-plane components of the nearest neighbor Fe-As force
constant matrix.
The ``original'' model failed to fit the data,
because (1) the in-plane polarized mode at $\Gamma$ point (A and B in
Fig.~\ref{Fig:dis_H00} with atomic motions A and B in
Fig.~\ref{Fig:motion}) has an energy that is too high compared to the
data, and (2) as one increases $q$ along the $<$100$>$ direction, the
model predicts an anti-crossing with a $c$-axis polarized mode (C in
Fig.~\ref{Fig:dis_H00}, and the atomic motion C in
Fig.~\ref{Fig:motion}) that is not observed.
However, selectively reducing the in-plane components of the Fe-As force
constant matrix by 20 \% reduces the energy of the
in-plane modes to be equal or just below that of the $c$-axis polarized
mode, and avoids the anti-crossing, in good agreement with the measured
data.
This ``in-plane soft'' model [Fig.~\ref{Fig:motion2}(a)-2] is compared
with the experimental data in Fig.~\ref{Fig:dis_comp_model}(b), and, for
being a relatively simple modification, agrees well with the measured
dispersion.
The improvement is also clearly seen along ($3+q$~$3+q$~0) in
Figs.~\ref{Fig:dis_comp_model_HK0_330}(a) and (b).
In particular this direction shows the necessity of softening only the
in-plane components instead of the full force constant matrices as we
did in Ref.~\onlinecite{FukudaT08a}.

Our third modified model begins with the force constant matrices from
the ``magnetic'' first principles calculation.
To reduce the splitting between ferromagnetic modes (whose calculated
frequencies agree with the data) and the antiferromagnetic ones (which
are calculated to have a higher energy than observed), we reduce the
$<$1\={1}0$>$ components of force constant matrices between nearest
neighbor Fe atoms.
Since the force constant between nearest neighbor Fe atoms is much
smaller (less than 15\%) than that between nearest neighbor Fe and As,
we ``clip'' these bonds by setting the force constant matrix to zero as
shown in Fig.~\ref{Fig:motion2}(b).
This ``clipped'' model is compared with the experimental data in
Fig.~\ref{Fig:dis_comp_model}(c), and the agreement is also relatively
good.

\begin{figure}
 \includegraphics[keepaspectratio,width=0.5\textwidth]{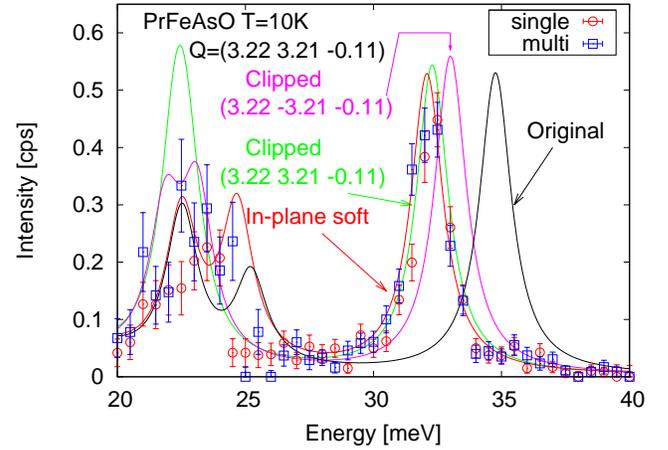}%
 \caption{\label{Fig:domain} (Color online)
 IXS spectra at $Q$ = (3.22~3.22~-0.11) in parent PrFeAsO sample at 10 K
 around the energy transfer of 30 meV; magnification of
 Fig.~\ref{Fig:dis_exp_110}(b).
 The symbols are also the same as those in Fig.~\ref{Fig:dis_exp_110}.
 While the red open circles show the measurement at a mostly single
 domain region, blue open squares show the measurement at a multi domain
 region.
 Two modified models as well as the ``original'' first principles
 calculation are also plotted with lines.
}
\end{figure}

The ``in-plane soft'' and ``clipped'' models have reasonable, though not
perfect, agreement with the experimental data in other symmetry
directions.
For the dispersion along the ($3+q$~0~0) (Fig.~\ref{Fig:dis_comp_model})
and (0~$q$~9) (not shown) the ``clipped'' model is perhaps slightly
better, while for the (3~$q$~0) direction (not shown) the ``in-plane
soft'' model is better.
However, if we consider the splitting in the ($3+q$~$3+q$~0) direction
of Fig.~\ref{Fig:dis_comp_model_HK0_330}, we find that the ``clipped''
model continues to predict some splitting that is not observed
experimentally.
This is also seen in the comparison of the IXS spectra of the parent
sample below $T_N$ at single and multi domain regions with the
``original'' {\itshape ab initio} calculation and two modified models in
Fig.~\ref{Fig:domain}.

\section{Discussion}\label{Sec:discussion}

The discussion above, after extensive investigation of several samples
of PrFeAsO$_{1-y}$ ($y\sim$0, 0.1, and 0.3) and comparison with
{\itshape ab initio} calculations and the modified models, allows the
following conclusions:
\begin{enumerate}
 \item There are small changes in the observed phonon intensities and
       dispersion when temperature or doping is
       modified.\label{item:unchange}
 \item The observed changes are generally much smaller than the rather
       large differences between the observed data and the {\itshape ab
       initio} calculations.\label{item:softening}
 \item Magnetic (spin-polarized LSDA) calculations tend to give better
       agreement with the measured data, because ferromagnetically
       polarized modes are softened.
       However, antiferromagnetically polarized modes are calculated to
       have energies that remain high and do not agree with
       measurement.\label{item:magnetism}
 \item Better agreement with the data can be obtained by modifying the
       calculation either by clipping the bond between
       antiferromagnetically polarized iron atoms in spin-polarized
       calculations, or by softening only the in-plane components of the
       force constant matrix of the Fe-As bond in nonmagnetic
       calculation.\label{item:modification}
\end{enumerate}
We now discuss these results in the context of other work.
At the start we note that variations of
\ref{item:unchange}--\ref{item:magnetism} have already been reported for
other iron-arsenide compounds.
Our work then demonstrates the general applicability of these
conclusions across many families of iron-arsenide compounds.
We also have put some results on rather firmer footing (e.g. the lack of
splitting of modes in explicitly twinned material).
Moreover, after a precise comparison with ``magnetic'' first principles
calculations, which had been previously considered to show good
agreement with experimental data by others, we still find significant
differences.
Consideration of these differences leads to two modified models
summarized in the above point \ref{item:modification}.

The presence of only small changes in phonon spectra resulting
from either temperature or doping, is in agreement with work by many
other authors.
Phonon DOS measurements by either INS or IXS have been carried out in
various iron-arsenide compounds; ``1111'' materials
[LaFeAs(O,F) (Refs.~\onlinecite{QiuY08a,ChristiansonAD08a,FukudaT08a}),
PrFeAsO$_{1-y}$ (Ref.~\onlinecite{FukudaT08a}), NdFeAs(O,F)
(Ref.~\onlinecite{TaconML08a}), Ca(Fe,Co)AsF
(Ref.~\onlinecite{MittalR09b}), and SrFeAsF
(Ref.~\onlinecite{ZbiriM10a})], ``122'' materials [(Ba,K)Fe$_2$As$_2$
(Ref.~\onlinecite{MittalR08a}), (Ca,Na)Fe$_2$As$_2$
(Ref.~\onlinecite{MittalR08b}), Ca(Fe,Co)$_2$As$_2$
(Ref.~\onlinecite{MittalR09c}), and SrFe$_2$As$_2$
(Ref.~\onlinecite{ZbiriM10a})], and ``11'' system [FeSe$_{1-x}$
(Ref.~\onlinecite{PhelanD09a})].
Fe partial phonon DOS measurements by nuclear resonant inelastic
scattering of synchrotron radiation is adapted to LaFeAs(O,F)
(Ref.~\onlinecite{HigashitaniguchiS08a}), (Ba,K)Fe$_2$As$_2$
(Ref.~\onlinecite{TsutsuiS10a}), Ba(Fe,Co)As$_2$
(Ref.~\onlinecite{DelaireO10a}), and Fe$_{1+\delta}$Se
(Ref.~\onlinecite{KsenofontovV10a}).
Using single crystals precise phonon dispersions have been investigated
by either IXS or INS in PrFeAsO$_{1-y}$ (Ref.~\onlinecite{FukudaT08a}),
(Ba,K)Fe$_2$As$_2$ (Ref.~\onlinecite{ReznikD09a}), and SmFeAs(O,F)
(Ref.~\onlinecite{TaconML09a}).
All of them found only rather small changes, if any, with doping and
temperature.
The exception is the change in (Ba,K)Fe$_2$As$_2$
(Ref.~\onlinecite{LeeCH10a}), but it is observed in only one particular
phonon mode.
Though (Ca,Na)Fe$_2$As$_2$ also shows some changes in phonon structure
with pressure\cite{MittalR09a}, this system is a relatively soft
material with a structural transition to ``collapsed tetragonal''
phase.
Raman\cite{LitvinchukAP08a,ZhaoSC09a,GallaisY08a,ZhangL09a,RahlenbeckM09a,ChauviereL09a,ChoiKY08a,ChoiKY10a,GranathM09a,KumarP10a,XiaTL09a,OkazakiK11a,ZhangAM11a,SugaiS10a}
and infrared (IR)
spectroscopy\cite{MirzaeiSI08a,AkrapA09a,WuD09b,HeumenE10a}, or
femtosecond-resolved pump-probe
reflectivity\cite{MansartB09a,MerteljT10a,TakahashiH11a}, are also used
to investigate phonons in these materials, and no significant change is
observed.
Our result of a generally weak dependence of the phonon spectra on
doping and temperature is then reasonably consistent with previous
work.
Higher resolution work does show some changes typically at the level of
0.5 meV.

Compared to the small temperature/doping dependences, the discrepancies
between the measured data and the {\itshape ab initio} calculations are
relatively large (i.e., a softening of observed Fe modes).
Moreover, the discrepancies become smaller if magnetism is included in
the calculations.
In general, this is also in good agreement with previous results.
Many phonon DOS as well as phonon dispersion measurements show the
softening of the experimental data compared with nonmagnetic {\itshape
ab initio} calculation by several meV on selected Fe and As phonon
modes\cite{ChristiansonAD08a,FukudaT08a,TaconML08a,ReznikD08a,HahnSE09a,ReznikD09a,TaconML09a,TaconML10a,ZbiriM09a,ZbiriM10a,MittalR09b}.
Among them some report the improvement possible using magnetic
calculations; phonon DOS of BaFe$_2$As$_2$
(Refs.~\onlinecite{YildirimT09b,ZbiriM09a}), $Re$Fe$_2$As$_2$ and
$Re$FeAsF (Ref.~\onlinecite{ZbiriM10a}), phonon dispersion of
CaFe$_2$As$_2$ (Ref.~\onlinecite{HahnSE09a}), SmFeAs(O,F)
(Ref.~\onlinecite{TaconML09a}), (Ba,K)Fe$_2$As$_2$ and
Ba(Fe,Co)$_2$As$_2$ (Refs.~\onlinecite{ReznikD08a,ReznikD09a}).
Here we note that the most pronounced softening is observed for high
energy plane-polarized modes ($\Delta E\sim$30--35 meV) in our present
work, while most of the previous results were focused to $\Delta
E\sim$20 meV especially in the phonon dispersions.

Given the generally good agreement of our calculations of the
ferromagnetically polarized modes with the measurements on the
magnetically ordered parent below $T_N$, the remaining discrepancy
between the calculated high energy for the antiferromagnetically
polarized modes and the measurements which show a lower energy
(essentially the same as the ferromagnetically polarized modes) is
surprising.
This is a strong result in that essentially all of our measurements only
show the lower energy modes, and is highlighted by the complete lack of
splitting (to the 0.5 meV level) for the occasion when we specifically
examined a twinned portion of the sample.
A similar lack of splitting was also observed by Reznik {\itshape et
al.}\cite{ReznikD09a} for BaFe$_2$As$_2$, though without explicit
confirmation of twinning.

We tried to understand the disagreement of data and calculations by ad
hoc modifications to our calculation results, and arrived at two models,
which have a completely different approach, but both of which show
relatively good agreement with the observed data.

First we consider the ``clipped'' model.
In this model we use the ``magnetic'' calculation, and include
additional in-plane anisotropy to cancel the anisotropy originating from
the magnetic order.
The observed in-plane phonon dispersion of even antiferromagnetically
ordered phase is surprisingly isotropic in the Fe-As plane.
Therefore, additional anisotropy in the force constant matrices is
indispensable, if we based it on ``magnetic'' calculation which it is
now widely believed to be the reasonable calculation of the
iron-arsenide system.
The anisotropy in the Fe-As plane, included in ``clipped'' model, is
also suggested by other experiments.
For example, from magnetic excitation measurements by INS large
anisotropic exchange coupling is suggested in CaFe$_2$As$_2$
[$SJ_a=49.9\pm9.9$ and $SJ_b=-5.7\pm4.5$ (Ref.~\onlinecite{ZhaoJ09a}) or
$24<J_a<37$ and $7<J_b<20$ (Ref.~\onlinecite{DialloSO09a})] or in
FeSe$_{0.5}$Te$_{0.5}$ (Ref.~\onlinecite{LeeSH10a}).
For these the strong correlation is along the ferromagnetic direction in
the ``clipped'' model [Fig.~\ref{Fig:motion2}(b)], while the spin-spin
coupling is stronger in the antiferromagnetic direction
[Fig.~\ref{Fig:motion2}(c)].
Moreover, recently orbital ordering in iron-pnictides has been
discussed.
Using the polarized laser angle-resolved photoemission spectroscopy
(ARPES) two-fold symmetry is observed in electronic structure of the Fe
3$d_{xz}$ orbital in BaFe$_2$As$_2$ below $T_N$
(Ref.~\onlinecite{ShimojimaT10a}).
Such orbital ordering also causes in-plane anisotropy and even might
explain\cite{LvW10a} the above-mentioned in-plane magnetic anisotropy of
the spin waves.
These anisotropic effects seem to suggest some support for the
``clipped'' model.
However, the residual anisotropy of mode energies in the ``clipped''
model is larger than observed, and attempts to reduce it to the level
measured (e.g., by including an attractive term) are not successful.
Given the evidences for anisotropy (e.g., in spin
wave\cite{ZhaoJ09a,DialloSO09a}, transport measurements\cite{ChuJH10a},
small ($\sim$0.5 meV) splitting of phonon modes in Raman
scattering\cite{ChauviereL09a}) it seems probable that the
reality is a nearly isotropic model, with some very small anisotropy.
However, the reason the direct {\itshape ab initio} calculations so
strongly over-estimate the anisotropy is not clear.

The ``in-plane soft'' model is an alternative based on the ``original''
microscopically in-plane isotropic calculation.
We slightly soften only the in-plane components of force constants while
preserving the out-of-plane component.
This is an improved version of our previous ``soft Fe-As''
model\cite{FukudaT08a}, based on the {\itshape nonmagnetic}
calculation.
However, it shows as good as or even better agreement with the measured
phonon structure compared to the ``clipped'' model.
Therefore, it seems worth considering nonmagnetic calculations again,
because experimentally the magnetic order does not strongly affect the
phonon structure, and the dispersion is essentially isotropic in many
iron-arsenides.

Finally we consider the effect of magnetic fluctuations, which is widely
discussed and has the possibility to reduce the in-plane anisotropy.
One model by Mazin and Johannes\cite{MazinII09d}, suggested there are
always antiferromagnetic domains, but the boundaries fluctuate.
Thus depending on the time scale of an experimental probe relative to
these fluctuations, different determinations may be made about the
presence of magnetism.
They suggested that many of the physical properties of iron-arsenide
compounds, both experimental and theoretical, may be explained by this.
With this model the small observed Fe magnetic moment is explained as by
averaging over zero-point and fast magnetic domain motion.
Thus, M{\"o}ssbauer experiments\cite{KitaoS08a,HigashitaniguchiS08a}
and $\mu$SR\cite{LuetkensH08a,CarloJP09a,TakeshitaS09a,OhishiK11a}, that
probe relatively slow ($>$10 ns) timescales would see primarily
nonmagnetic response, as they do, while probes of faster time scales
might see magnetic response, such as the splitting observed in
photo-emission\cite{BondinoF08a}.
INS investigations, in fact, show magnetic fluctuations at finite energy
($\sim$0.2 ps time scales) even in superconducting samples and/or at
high temperature above
transition\cite{ChristiansonAD08b,IshikadoM09a,WakimotoS10a}.

However, it seems unlikely to us that the presence of
fluctuations can fully explain the difference between our calculations
and our data.
Phonon time scales are given by the oscillation period ($\sim$0.1 ps for
a 30 meV mode), and the phonon lifetime ($\sim$0.7 ps for a 1 meV mode)
linewidth.
Fluctuations on that are slow on the ps time scale should lead to
splitting of modes, while those with timescales $<$0.1 ps should lead to
unsplit phonons that see the average structure.
The magnetic calculations give mode splittings of several meV while for
the particular case of Fig.~\ref{Fig:dis_H00}, for example, our data
would support splittings no larger than a few tenths of a meV.
Also, the observed mode energy is in good agreement with the one of the
two calculated modes, not their average.
Thus, we speculate there is some additional neglected factor, not only
magnetic fluctuations, that leads to the overestimation of the
calculated phonon anisotropy in these systems.

\section{Conclusion}\label{Sec:conclusion}

Extensive investigation of phonon dispersion in several samples of
PrFeAsO$_{1-y}$ allow careful comparison of results between different
samples, and with calculations.
Small changes were visible as the material was doped into the
superconducting state (from $y=0$ to $y\sim 0.3$), and when the parent
materials was cooled below the magnetic and structural phase transition.
Smaller changes were visible when superconducting samples were cooled
across $T_c$.
Interpretation of the observed small changes with doping or temperature
in this system is complicated by the generally larger disagreement with
calculations.
Magnetic calculations compare more favorably to the data, especially the
ferromagnetically polarized phonon modes, but show splitting between
ferromagnetically and antiferromagnetically polarized modes (several
meV) that is not at all supported by the experimental data ($<$0.5
meV).
Interestingly, the in-plane isotropy of phonon properties can be
constructed from both microscopically anisotropic (magnetic) and
isotropic (nonmagnetic) calculations.
It is clear that while the {\itshape ab initio} calculations provide a
reasonable rough estimate of phonon dispersion, they lack some
fundamental ingredient.
Given the similarity of the measured response for both the magnetically
ordered parent material below $T_N$ and the other systems, however, is
seems unlikely that the missing ingredient might be only magnetic
fluctuations.

\begin{acknowledgments}
 Work at SPring-8 was carried out under proposal No. 2008A2050,
 No. 2008B1403, No. 2009A1436, No. 2009B1609, No. 2009B2136, and
 No. 2010A1296.
 This work was performed under the NIMS-RIKEN-JAEA Cooperative Research
 Program on Quantum Beam Science and Technology.
\end{acknowledgments}


\end{document}